\documentclass[11pt,twoside]{article}
\def\grtsim{\mathrel{\hbox{\rlap{\hbox{\lower2pt\hbox{$\sim$}}}\raise2pt\hbox{$>$}}}}
\newcommand{\gtsim}{\mbox{{\raisebox{-0.4ex}{$\stackrel{>}{{\scriptstyle\sim}}$}}}}

\usepackage{galev06}
\usepackage{epsf}
\usepackage{psfig}
\usepackage{lscape}
\usepackage{graphicx}
\usepackage{lscape}

\begin{document}
\title{Probing accretion activity in radio sources using 24 $\mu \rm m$ Spitzer data}
\author{Eleni Vardoulaki$^{1}$, Steve Rawlings$^{1}$, Chris Simpson$^{2}$}
\affil{$^{1}$ 
Astrophysics, Department of Physics, Denys Wilkinson Building, Keble Road,
Oxford, OX1 3RH, UK\\
$^{2}$
Astrophysics Research Institute, Liverpool John Moores
University, Twelve Quays House, Egerton Wharf, Birkenhead CH41 1LD, UK
}

\begin{abstract} 
We study the 36 brightest radio sources in the Subaru/\textit{XMM-Newton} 
Deep Field (SXDF). Using MIPS 24 $\mu \rm m$ data from Spitzer we expect to 
trace accretion activity, even if it is hidden at optical wavelengths, unless 
the obscuring column is extreme. Our results suggest that in the decade or so 
below the break in the radio luminosity function that at least half, and 
potentially nearly all, radio sources are associated with accreting 
quasar-like objects. This is not true at lower radio luminosities where the 
quasar-like fraction approaches zero once compact sources are excluded.
\end{abstract}

\section{Introduction}

In unified AGN models, powerful radio sources are believed to have a 
central source emitting in the optical, UV and X-rays. The dusty torus that 
surrounds the nuclear region absorbs this light and re-emits it in the 
infrared. This mechanism makes it difficult to observe the objects viewed 
through the torus directly in the optical, UV and soft X-rays. Willott et 
al. (2000) define the quasar fraction $f_{q}$ to be the ratio of unobscured to 
obscured quasars. In \cite{eleniv06} we expressed the need to study radio 
sources in the mid-infrared in order to investigate the existence of any 
hidden  accretion activity in radio sources. Spitzer observations at 24 
$\mu \rm m$ are ideal for this task, since they trace hot dust emission 
which can be obscured only by extreme columns (\cite{ogle06}).

Our sample is the 36 brightest radio sources from the VLA survey of the SXDF 
(\cite{simpson06}) with flux densities greater than 2 mJy at 1.4 GHz. Most of 
our objects are not as yet spectroscopically confirmed, so we use photometric 
redshifts for our analysis. We use a radio spectral index $\alpha$ = 0.8 
($S_{\nu} \propto \nu^{-\alpha}$). We assume throughout a low-density, 
$\Lambda$-dominated Universe in which
$H_{0}=70~ {\rm km~s^{-1}Mpc^{-1}}$, $\Omega_{\rm M}=0.3$ and 
$\Omega_{\Lambda}=0.7$.

More details of this work will appear in Vardoulaki et al. (in prep.).

\section{Discussion}  

\begin{figure} 
\plottwo{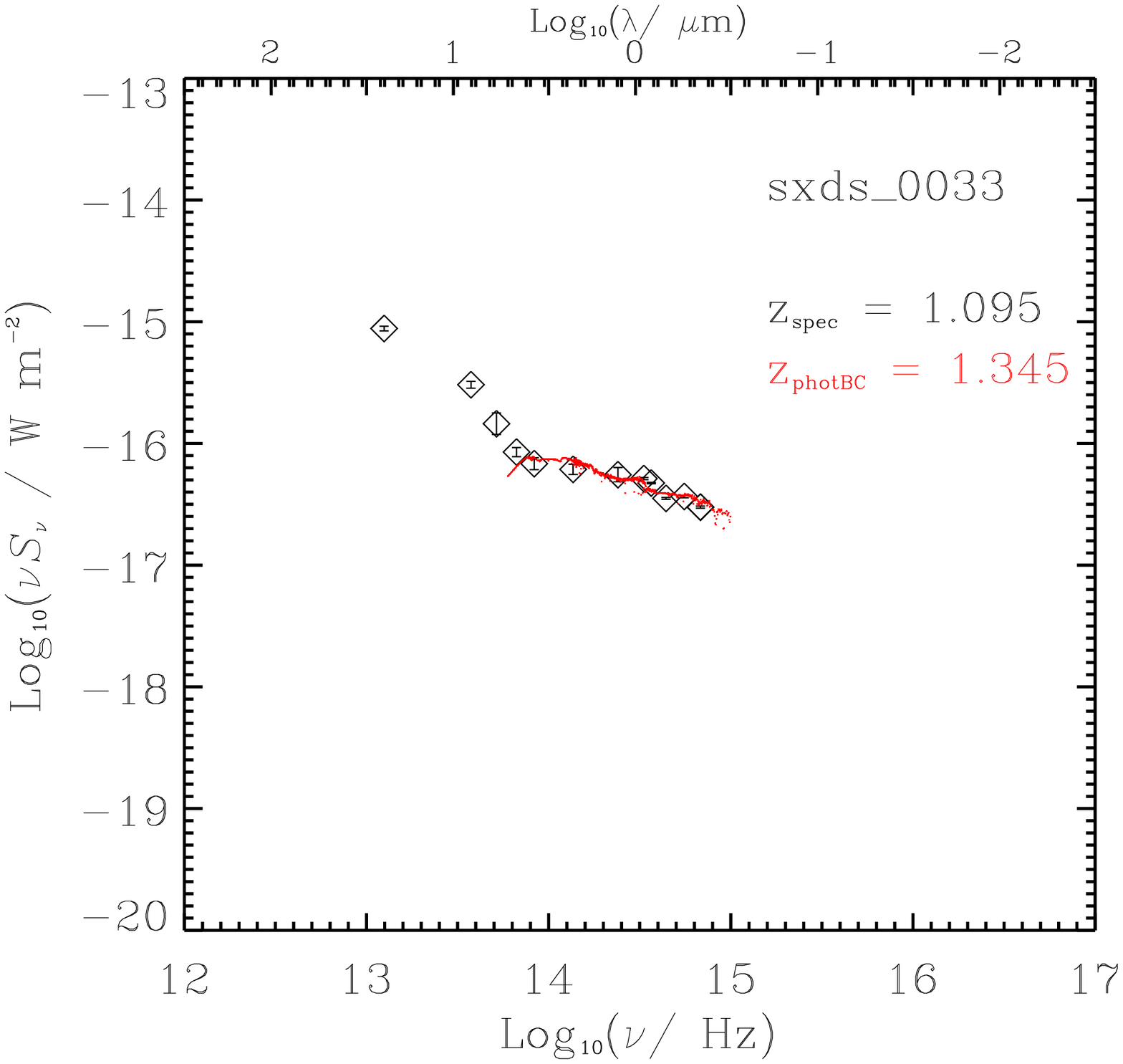}{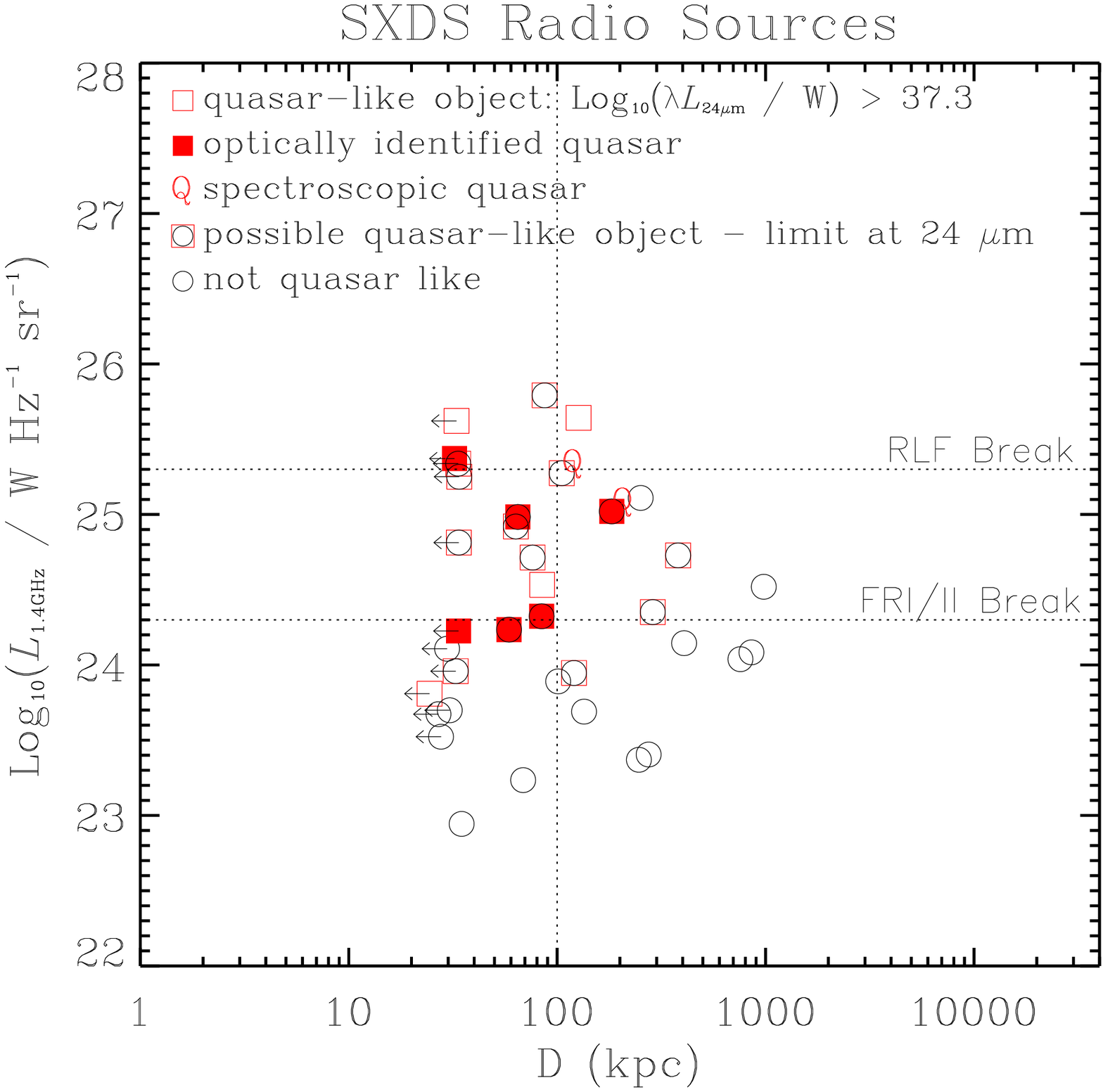}
\caption{{\itshape Left:\/} Observed frame SED of sxds\_0033 with a redshift 
estimated using HyperZ and data from Spitzer, UKIDSS and SXDF surveys. 
{\itshape Right:\/} $\log_{10}( L_{1.4{\rm GHz}})$ versus largest projected 
linear size $D$: possible quasar-like objects have upper limits on 
$\lambda L_{24 \mu \rm m}$ consistent with 
$\lambda L_{24 \mu \rm m}>10^{37.3}$ ${\rm W}$.}
\label{fig1}
\end{figure}

We have constructed spectral energy distributions (SEDs) for our SXDF sample 
and fitted Bruzual-Charlot galaxy templates using the HyperZ photometric code 
(\cite{hyperz}) to estimate photometric redshifts. An example is shown on 
the left side of Fig. \ref{fig1}. This object is spectroscopically confirmed 
at redshift $z$ = 1.095. From its SED we can see excess emission at 24 
$\mu \rm m$, indicating that this object has hidden accretion activity. This 
object is not a spectroscopically confirmed quasar but has high-excitation 
narrow lines in its optical spectrum; its radio map shows an unresolved source.

We define objects to be `quasar-like' when their rest-frame luminosity at 
$\lambda =$ 24 $\mu \rm m$ is $\lambda L_{24 \mu \rm m}>10^{37.3}$ ${\rm W}$, 
since this corresponds to  $\lambda L_{24 \mu \rm m} > 10^{-1.8} L_{Edd}$ for 
a quasar with $M_{\rm BH}$ $\gtsim$ $10^{8}$ $\rm M_{\odot}$ 
(see \cite{mclure04}; \cite{mclureetal04}). The luminosity at 24 $\mu \rm m$ 
is calculated using the spectral index measured from 24 $\mu \rm m$ to the 
nearest lower-$\lambda$ detection in the observed frame SED. The quasar-like 
fraction $q_{\rm l}$ in the region between the RLF break and the FRI/II break 
is high, $q_{\rm l} \sim 0.5-0.9$, depending on how many of the 24 
$\mu \rm m$ upper limits turn out to be close to detections. [The most radio 
luminous of the two non-quasar-like objects, is an FRI at $z \sim 0.7$]. 
Recall that $f_{q} \sim 0.2$ in this region (\cite{willott00}), i.e. the 
24 $\mu \rm m$ data have revealed at least some hidden accretion activity.

The quasar-like fraction drops significantly below the FRI/II break, and if 
one excludes compact ($D <$ 100 kpc) sources as potentially part of a separate 
(beamed) population, we see that there are no definite quasar-like objects.
%($q_{\rm l}$ $\ltsim$ 0.1).

We conclude that Spitzer 24-$\mu \rm m$ data on the SXDF radio sources are 
consistent with the hypothesis that quasar-like objects are almost always 
associated with powerful (above the FRI/II break) radio sources, but are 
rarely connected to less luminous sources, unless they are part of the compact 
population.

\end{document}